\documentclass[11pt]{article}
\usepackage[a4paper,top=3cm,bottom=3cm,left=2cm,right=2cm,marginparwidth=1.75cm]{geometry}
\usepackage{graphicx} % Required for inserting images
\usepackage{amsmath}
\usepackage{hyperref}
\hypersetup{
    colorlinks=true,
    linkcolor=blue,
    filecolor=blue,      
    urlcolor=blue,
    allcolors=blue,
    pdftitle={Overleaf Example},
    pdfpagemode=FullScreen,
    }

\usepackage{apacite}
\usepackage{authblk}
\usepackage{lineno}
\usepackage{multirow}
\usepackage{tabularx}
\usepackage{physics}

\usepackage{caption}
\title{\textbf{Warming from cold pools: A pathway for mesoscale organization to alter Earth's radiation budget}}

\author[1]{Pouriya Alinaghi\footnote{Correspondence 1: p.alinaghi@tudelft.nl}}
\author[1,2]{Martin Janssens\footnote{Correspondence 2: martin.janssens@wur.nl}}
\author[1]{Fredrik Jansson}

{\small
\affil[1]{Delft University of Technology, Delft, The Netherlands}
\affil[2]{Wageningen University \& Research, The Netherlands}
}

\date{Date: \today}

\begin{document}

\maketitle

\section*{Abstract}
Marine shallow cumulus clouds have long caused large uncertainty in climate projections. These clouds frequently organize into mesoscale (10-500 km) structures, through two processes that couple the clouds to shallow mesoscale circulations: (i) mesoscale moisture aggregation, and (ii) cold pools, driven locally from rain-evaporation. Since global climate models do not capture these mesoscale processes, while the degree of mesoscale organization is observed to correlate to shortwave cooling, it has been suggested that mesoscale processes modulate the cloud response to global warming. Here, we show that introducing mesoscale dynamics can indeed substantially alter top-of-the-atmosphere radiative budget, if the balance between the two circulations is upset. By homogenizing rain-evaporation across the horizontal domain, we suppress the cold-pool-driven circulations in a large ensemble of large-domain, large-eddy simulations. We find that cold pools reduce mesoscale ascent, thereby arresting a runaway self-aggregation of moisture into very moist regions. This reduces the net rainfall of the cumulus fields, moistens the boundary layer and thus reduces the emission of clear-sky longwave radiation to space, giving an ensemble-averaged warming of 1.88\,W/m$^2$. Our results highlight that the proper interplay between mesoscale processes is critical for capturing radiative budgets—especially in kilometer-scale climate models that only partially resolve aggregation and cold pools.

\newpage
\section{Introduction}
Uncertainties in the response of tropical shallow cumulus clouds to global warming are a long-standing source of uncertainty in climate projections \cite{bony2004dynamic,bony2005marine,vial2013interpretation,bony2015clouds,schneider2017climate,nuijens2019boundary,zelinka2020causes,sherwood2020assessment}. Such uncertainty is usually framed in terms of predictions of the ``shortwave cloud feedback'', the change in sunlight reflected back to space by clouds per unit warming of the tropical environment. Predictions of this feedback diverge among global climate models (GCMs) \cite{bony2004dynamic,bony2005marine,vial2013interpretation,schneider2017climate}, high-resolution atmospheric models \cite{blossey2013marine,radtke2020shallow} and observation-based estimates \cite{cesana2021observational,myers2021observational}, because the models inadequately represent several cloud processes which are thought to control shortwave cooling \cite{vogel2022strong,vial2023cloud,bellouin2020bounding,stier2024multifaceted}.

One missing process class comprises ``mesoscale processes", which act on horizontal scales of tens to hundreds of kilometers, to ubiquitously organize fields of shallow cumuli into mesoscale patterns \cite{stevens2020sugar,janssens2021cloud,denby2020discovering,rasp2020combining,bony2020sugar,schulz2021characterization,schulz2022}. GCMs discretize the atmosphere too coarsely to resolve these processes, while detailed process models have traditionally covered domains too small for the cumuli to organize. Mesoscale processes are hypothesized to influence cloud feedback, because the shortwave cooling of cumulus fields rises in proportion with their degree of mesoscale organization \cite{bony2020sugar,alinaghi2023shallow,denby2023charting}. Furthermore, in fields of \textit{deep} cumulus clouds, the cooling from outgoing clear-sky longwave radiation also rises with mesoscale organization \cite{tobin2012observational,muller2012detailed,muller2013aggregation,bony2020observed}. Hence, the same missing mesoscale processes that make GCMs' shortwave feedback uncertain in regions where shallow cumulus clouds prevail, could well alter their \textit{longwave} feedback, though this pathway remains unexplored.

Yet, even if organized shallow cumulus fields produce more radiative cooling to space than unorganized cloud fields, this does not necessarily mean that the unorganized fields simulated by climate models cannot respond correctly to warming. A necessary, but unproven, condition for mesoscale processes to alter contemporary estimates of cloud feedback, is that their inclusion in climate models meaningfully alters the top-of-atmosphere (TOA) radiation budget, given the same climatological forcing. 

Here, we test this condition through detailed atmospheric process models, large-eddy simulations (LES) \cite{heus2010formulation}, on domains large enough to simulate mesoscale organization (153.6$\times$153.6\,km$^2$). Under fixed and spatially homogeneous large-scale conditions, these LESs produce mesoscale organization through two pathways: First, non-precipitating cumulus clouds spontaneously self-aggregate into mesoscale clusters via shallow, self-reinforcing mesoscale circulations driven by condensational heating in the clouds \cite{bretherton2017understanding,narenpitak2021sugar,janssens2023nonprecipitating,janssens2024shallow}. Second, as the cumuli deepen and begin to precipitate, rain evaporation in the cloudy columns forms pockets of cold, dense air. While downdrafts within these ``cold pools" suppress cloud formation locally, their expanding fronts carry moist air and strong updrafts that trigger new cumuli, forming ring-like mesoscale cloud structures \cite{zuidema2012trade,seifert2013large,li2014simulated,vogel2016role,zuidema2017survey,helfer2021morphology,lamaakel2022organization,lamaakel2023computational,alinaghi2024external,alinaghi2024cold}. In both pathways, shallow cumulus clouds organize through mesoscale circulations, a coupling also recently observed \cite{bony2017eurec,stevens2021eurec,george2023widespread,vogel2022strong}. Yet, the two circulations are formed by distinct processes: moisture self-aggregation and rain evaporation. When both processes are active in large-domain LESs, they do not appear to affect the TOA radiative budgets \cite{janssens2024symmetry}.

However, in this work, we show that this result emerged from a balance of individual mesoscale processes which \textit{can} affect the climatological cooling of fields of shallow cumuli. To demonstrate this, we compare simulations with both mesoscale processes active, to simulations where we prevent the cold-pool-driven circulations from forming: Instead of allowing rain-evaporation to locally cool the atmosphere and make it negatively buoyant, we redistribute this cooling evenly across a horizontal model level, such that no negative buoyancy perturbations are created \cite{boing2012influence}. To assess the significance of this cold-pool-denial's effect on the top-of-atmosphere radiation balance, we compare its magnitude to the variability in TOA radiation induced by varying ``cloud controlling factors'' (CCFs) \cite{stevens_2009_lowccf,klein_2018_ccf,myers2013observational,scott2020observed}, aspects of the large-scale environment whose modifications control the cloud response to warming in GCMs \cite{golaz2002pdfbased,hourdin2019unified,walters2019metoffice}. The denial experiment is therefore conducted in two otherwise identical 19-member, large-domain LES ensembles, in which we vary four CCFs to which cold pools are sensitive: large-scale geostrophic wind speed $\vert u_0 \vert$, subsidence $w_1$, stability $\Gamma$ and wind shear $u_z$ \cite{alinaghi2024external,alinaghi2024cold} (Adding other CCFs to which cold pools are insensitive, e.g. free-tropospheric humidity, would not affect cold-pool characteristics or alter the estimated radiative impact of suppressing cold-pool formation; see \textit{Methods}). We name these ensembles ``CP'' (cold pools are active) and ``NoCP'' (cold pools are homogenized).

\begin{figure}[t!]
\centering
\includegraphics[width=0.8\linewidth]{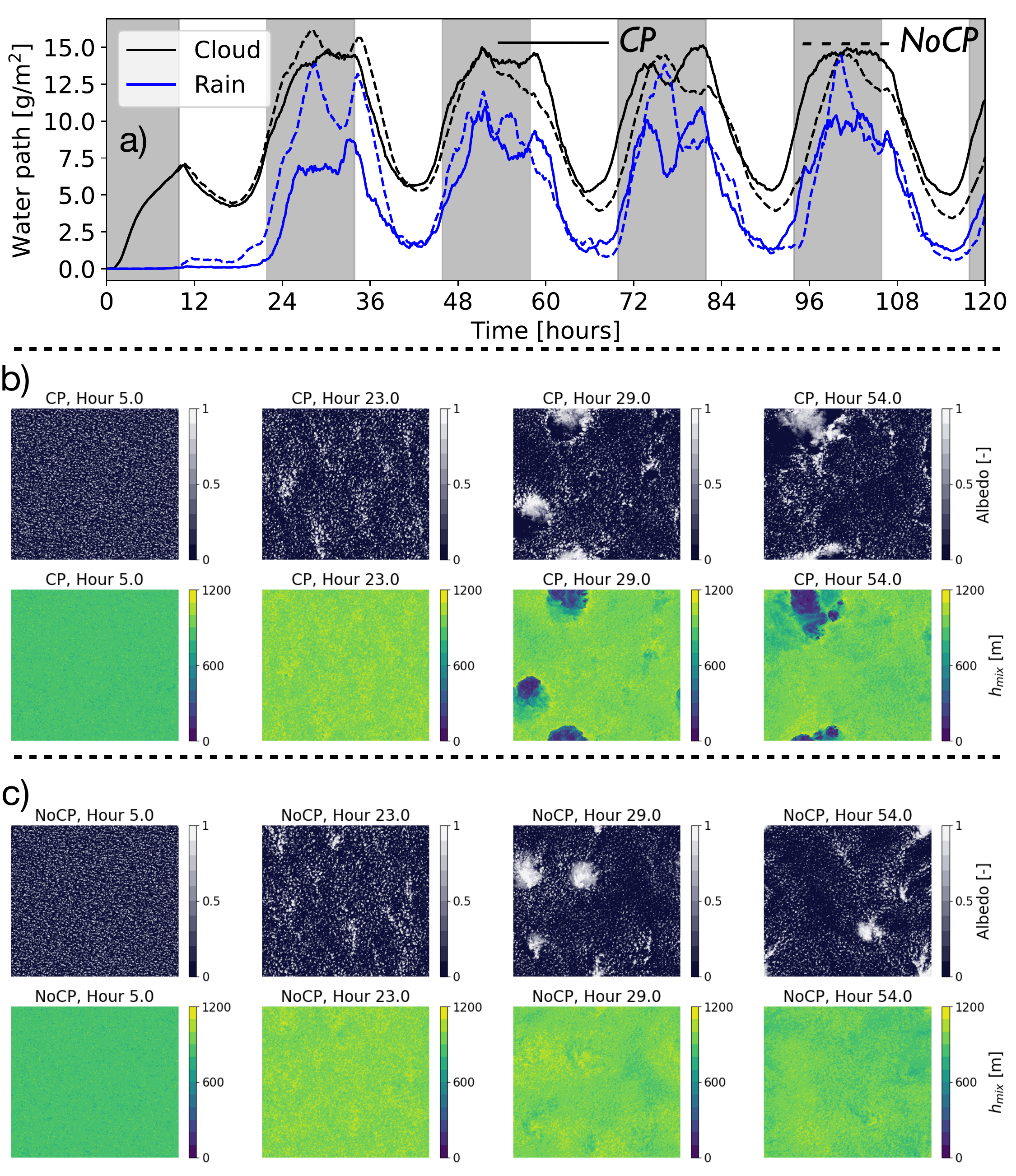}
\caption{{\textbf{Temporal evolution of cloud fields in both the CP and NoCP ensembles.}} a) Time series of cloud- and rain-water paths averaged across the entire CP (continuous) and NoCP (dashed) ensembles. Grey shades show nighttime. b) and c) are the top-view fields of cloud albedo (first row) and mixed-layer height (second row) for the central reference simulation of the CP and NoCP ensembles, respectively.}\label{fig: cloud fields}
\end{figure}

Both CP and NoCP ensembles are run for five daily cycles of insolation, starting from identical, cloud-free initial conditions. Initially, both evolve similarly as non-precipitating cumuli self-aggregate via shallow circulations (Fig. \ref{fig: cloud fields}a-c). As clouds deepen and precipitation begins, cold pools emerge in the CP ensemble, forming mesoscale ring-like cloud structures enclosing clear-sky areas and suppressing the height of the near-surface well-mixed layer $h_{mix}$ (Fig. \ref{fig: cloud fields}b), consistent with prior modeling and observational studies \cite{rochetin2021physically,touze2022cold}. In contrast, the NoCP ensemble precipitates without forming cold pools; $h_{mix}$ remains elevated, and the clouds continue to self-aggregate without generating open-sky regions (Fig. \ref{fig: cloud fields}c).

We trace the impact of cold pools on the TOA radiation budget in three parts. First, we quantify the difference in net TOA radiation between the CP and NoCP ensembles, showing that cold pools provide a significant warming at the TOA, not by modifying the fraction of reflected sunlight, but by reducing the outgoing longwave clear-sky radiation. This reduced longwave cooling originates in the CP ensemble's more humid boundary layers. Thus, in the second part, we show that cold pools keep the boundary layer humid by producing less rainfall. Third, we demonstrate that cold pools reduce the surface rainfall, because they counteract the circulations which aggregate water vapor into very moist regions in their absence, thus preventing efficient rainfall in these regions. Hence, we suggest the key to understanding whether mesoscale processes alter the cloud feedback, lies in understanding the interplay between shallow mesoscale circulations and the field of low-lying water vapor.

\section{Results}
\label{sec: results}
% \subsection*{Inhibiting cold-pool formation governs more clear-sky longwave radiative cooling}
\subsection*{Cold pools warm by reducing clear-sky outgoing longwave radiation}
\label{sec: 1}

\begin{figure}[t!]
\centering
\includegraphics[width=0.8\linewidth]{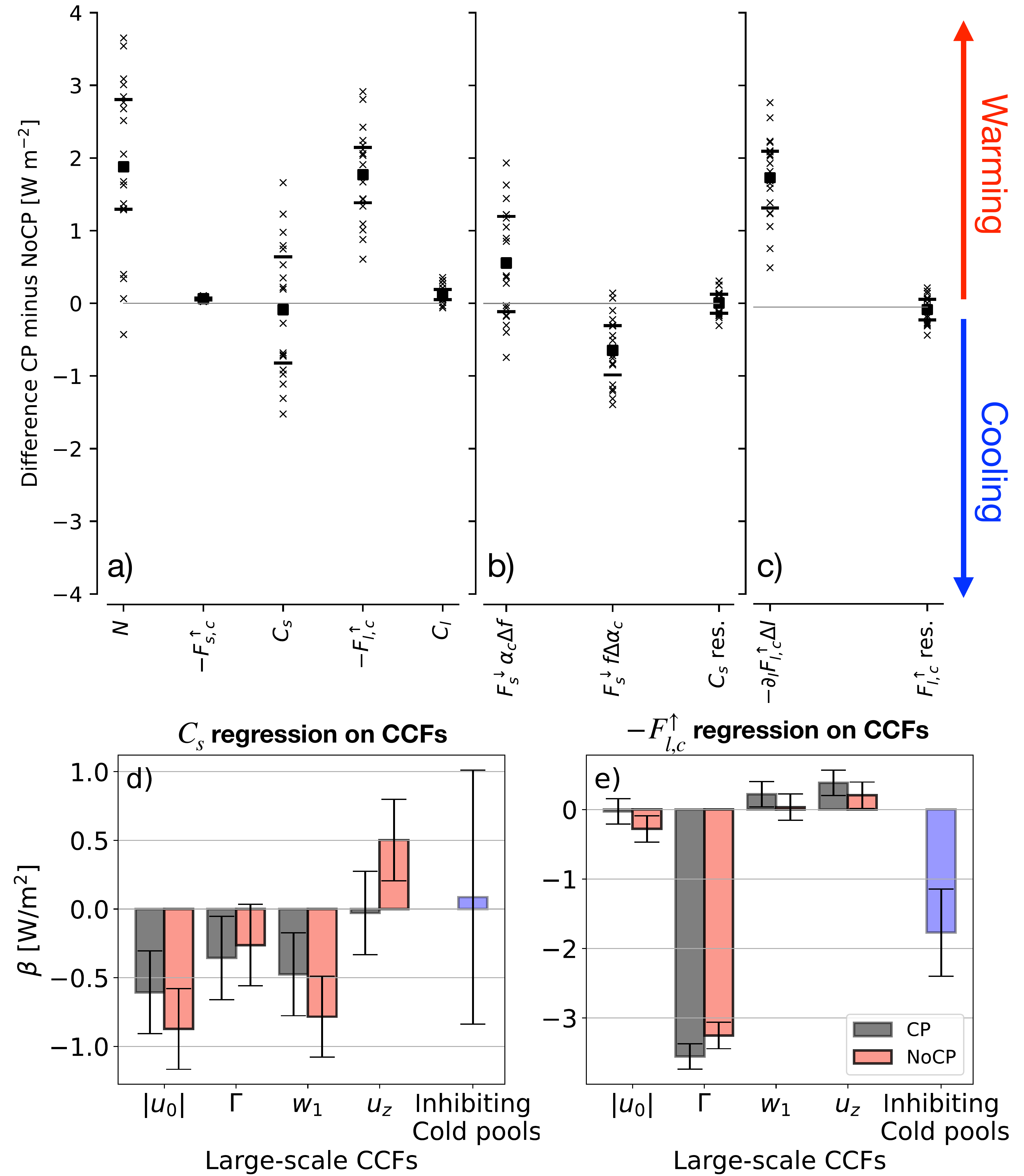}
    \caption{\textbf{Effects of cold pools on the daily mean net radiative budget at the top of the atmosphere.} a) Decomposition of the net radiative flux at the TOA into shortwave $C_s$ and longwave $C_l$ cloud radiative effects, and the shortwave $F_{s,c}^\uparrow$ and longwave $F_{l,c}^\uparrow$ outgoing clear-sky fluxes at the TOA. For each component, the y-axis shows the associated TOA radiation of CP ensemble minus that of NoCP ensemble. All fluxes are averaged over the last (5$^{th}$) day of each ensembles' member. Ensemble member means are shown as crosses ($\times$), ensemble means as squares, interquartile ranges as horizontal bars. b) Decomposition of $C_s$ to effects from cloud fraction $F^{\downarrow}_s\alpha_c\Delta f$ and albedo $F^{\downarrow}_sf\Delta \alpha_c$. c) Dependence of $\Delta F_{l,c}^\uparrow$ on the difference between the mean total-water path $\Delta I$ between the CP and NoCP ensembles. d,e) Beta coefficients (sensitivities) of the multiple regression of $C_s$ (R$^2 = 0.74$ for the CP and R$^2 = 0.88$ for the NoCP ensemble) and $-F_{l,c}^\uparrow$ (R$^2 = 0.99$ for both CP and NoCP ensembles) on CCFs (Note: the larger $w_1$, the weaker the large-scale subsidence in the ensemble). The p-values of the F-statistic of the model are smaller than $10^{-12}$. The impact of inhibiting cold-pool formation (NoCP$-$CP) is compared with the impact of the unit of change in CCFs.}
\label{fig: net radiation decomposition}
\end{figure}

The net TOA radiative budget $N$ is expressed as $N = C_s + C_l - F_{s,c}^\uparrow - F_{l,c}^\uparrow$, where $C_s$ and $C_l$ represent the shortwave and longwave cloud-radiative effects, respectively, while $F_{s,c}^\uparrow$ and $F_{l,c}^\uparrow$ are the outgoing shortwave and longwave clear-sky fluxes. A positive $N$ indicates energy accumulation and warming. All terms are domain- and time-averaged on day five for each CP and NoCP ensemble member, when the simulated cloud layers no longer change much. The mean TOA net radiative budget $N$ for the CP ensemble exceeds that of the NoCP ensemble by 1.88 W/m$^2$ (Fig.  \ref{fig: net radiation decomposition}a). The relative warming of the CP ensemble ($N_\text{CP} - N_\text{NoCP} > 0$) is robust for 18 out of 19 ensemble members. This warming is mainly due to reduced $-F_{l,c}^\uparrow$ in the CP ensemble, as differences in $C_s$, $C_l$, and $F_{s,c}^\uparrow$ are negligible.

The muted response of $C_s$ to homogenizing rain evaporation is composed of offsetting effects (Fig. \ref{fig: net radiation decomposition}b): reduced cooling from smaller cloud fraction ($F^{\downarrow}_s\alpha_c\Delta f$) is balanced by enhanced cooling from higher albedo in the CP ensemble ($F^{\downarrow}_sf\Delta \alpha_c$), where $F^{\downarrow}_s$ is the insolation flux. A $ 1.5$ W/m$^2$ spread around $\Delta C_s$ arises from the CP ensemble's higher albedo and lower coverage, with their relative impact differing across the ensemble (adding more ensemble members with diverse CCFs would yield a more representative estimate of cold-pool impacts on $C_s$; see \textit{Methods}). A similar compensation explains the minimal difference in $C_l$: Clouds in the CP ensemble are smaller in horizontal extent and fraction $f$, but deeper. Both more and deeper clouds enhance longwave warming, and these effects offset.

Thus, the net warming from cold pools stems from decreased $-\Delta F_{l,c}^\uparrow$ (Fig. \ref{fig: net radiation decomposition}c). This can be attributed to the difference in the horizontally and temporally averaged vertically-integrated total water (cloud + vapor) $I$ between these CP and NoCP ensembles, $\Delta F_{l,c}^\uparrow \approx \partial_I F_{l,c}^\uparrow \Delta I$. Since $I$ is mostly composed of water vapor concentrated in the boundary layer (see \textit{Methods}), our cold pools warm the simulation domains by keeping the boundary layer moist, which prevents the emission of longwave radiation to space  through clear skies \cite{bony2020observed,fildier2023moiture}.

The cold-pool-induced warming is significant: Multivariate regression of $-F_{l,c}^\uparrow$ onto the CCFs shows that cold pools have a greater impact on $-F_{l,c}^\uparrow$ compared with a unit of change in $\lvert u_0 \rvert$, $w_1$, or $u_z$ (Fig. \ref{fig: net radiation decomposition}e). Only enhancing the free-tropospheric stability $\Gamma$ has a larger impact on $-F_{l,c}^\uparrow$ than cold pools. The much weaker, uncertain impact of cold pools compared to CCFs on $C_s$ is shown for reference (Fig. \ref{fig: net radiation decomposition}d). Furthermore, the 1.88\,W/m² of biased cooling at the TOA—induced by cold-pool suppression—in the current climate has the potential to influence the longwave clear-sky feedback, given recent findings that emphasize the importance of the contemporary mean state of the tropical marine low clouds in shaping their response to warming \cite{ceppi2024implications}.

\subsection*{Cold pools lead to lower drying rates due to weaker precipitation}
\label{sec: 2}

\begin{figure}[t!]
\centering
\includegraphics[width =0.8\linewidth]{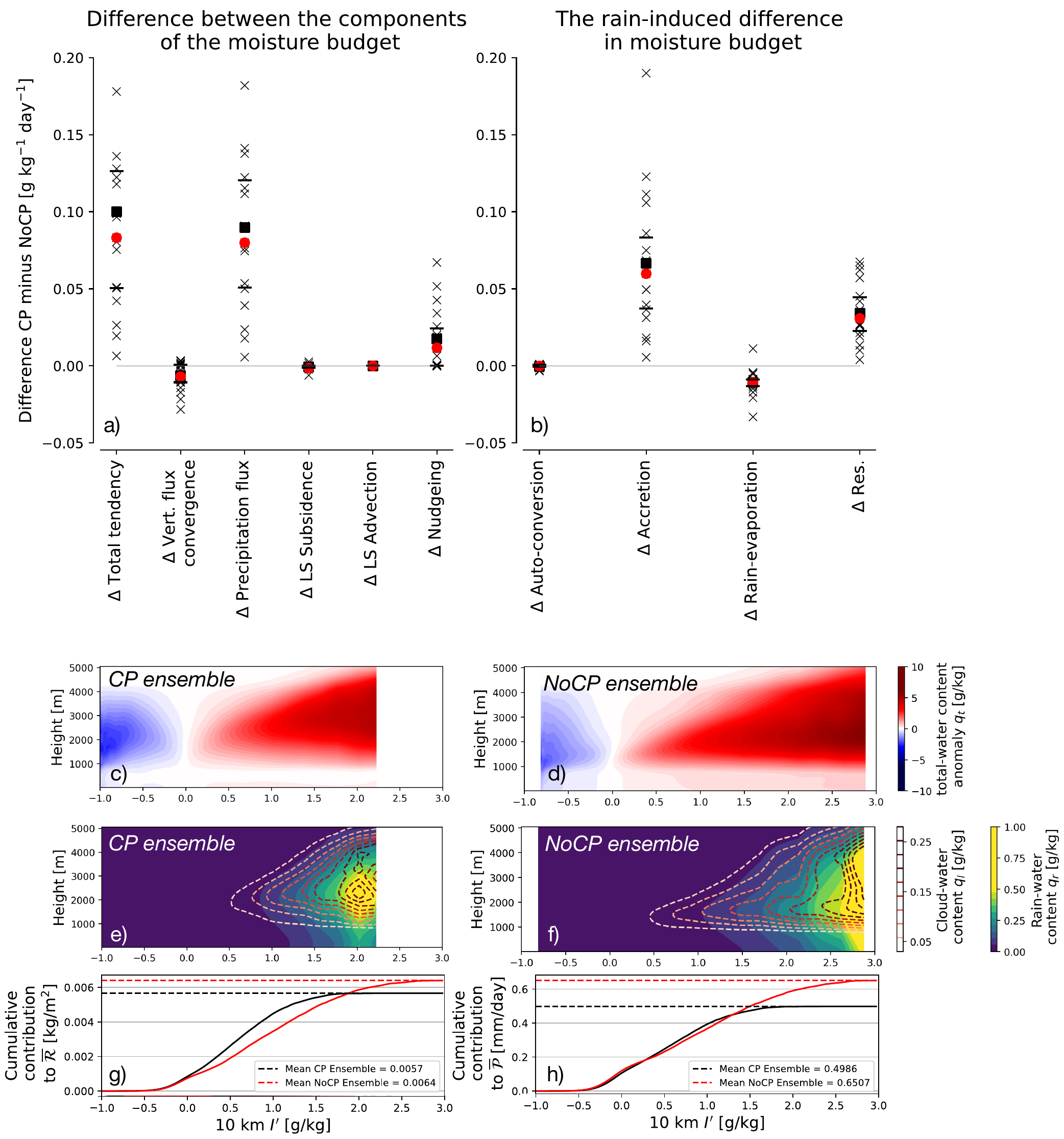}
\caption{\textbf{The impact of cold pools on the bulk moisture budget through precipitation flux.} Decomposition of (a) the bulk moisture budget and (b) the rain-induced difference in moisture budget across ensemble, averaged over hours 12–36 when simulations begin diverging in total moisture. Ensemble member means are shown as crosses ($\times$), ensemble means as squares, interquartile ranges as horizontal bars, and the central reference simulation as a red circle (see Fig. S1 for its temporal evolution). Moisture space of the total-water content anomaly $q_t'$ (c–f) Moisture space of total-water anomaly $q_t'$ and cloud $q_l$ rain $q_r$ water contents for CP and NoCP ensembles; $q_l$ is indicated by dashed white-to-red contours. (g–h) Cumulative contributions of $I'$ bins to ensemble-mean $\mathcal{R}$ and $\mathcal{P}$ for CP (black) and NoCP (red) ensembles.}
\label{fig: bulk moisture and precip}
\end{figure}

To understand why CP simulations remain moister than their NoCP counterparts, we evaluate the ``bulk" (vertically integrated) moisture budget in our simulations' lower atmospheres (see \textit{Methods}). These attribute the relative moistening of the CP ensemble's trade-wind layers to a reduction in drying by rainfall (precipitation flux, fig. \ref{fig: bulk moisture and precip}a). In turn, the smaller precipitation fluxes in the CP ensemble are due to lower rates of rain production by accretion (Fig. \ref{fig: bulk moisture and precip}b). In our LES's microphysics scheme, accretion rates are proportional to the product of cloud- and rain-water content, $q_l$ and $q_r$ \cite{seifert2001double}, so we next investigate their differences between the CP and NoCP ensembles.

For rain formation, clouds must deepen. In the trade-cumulus regime, cloud deepening is correlated with horizontal widening \cite{alinaghi2023shallow,feingold2017analysis}. This scale growth is driven by shallow, self-reinforcing mesoscale circulations driven by condensational heating over anomalously moist regions \cite{bretherton2017understanding,narenpitak2021sugar,janssens2023nonprecipitating}. We therefore analyze how $q_l$ and $q_r$ relate to the spatial variability in column-integrated total moisture, $I'$, in both ensembles. To this end, we divide the CP and NoCP ensembles into 10-km mesoscale blocks and compute the block-averaged $I$, denoted $I_m$. The mesoscale moisture anomaly is then defined as $I' = I_m - \overline{I}$, where $\overline{I}$ is the domain-mean $I$. Blocks are grouped into bins of ascending $I'$, and for each bin, we compute the average $q_l$, $q_r$, and the total-water content anomaly ($q_t'$).

These ``moisture-space'' plots illustrate that $q_t'$ is consistently positive at all heights where $I' > 0$, and negative where $I' < 0$ (Fig. \ref{fig: bulk moisture and precip}c,d). The mesoscale moisture anomalies are concentrated in the upper cloud layers, where they can significantly influence raindrop growth. Indeed, blocks with higher $I'$ host deeper clouds with greater $q_l$ and $q_r$ on top of the elevated $q_t$ anomalies (Fig. \ref{fig: bulk moisture and precip}e,f), where accretion is most efficient at producing rain. However, the CP ensemble lacks the extremely moist mesoscale regions with $2<I'<3$ g/kg that are present in the NoCP ensemble. It is the presence of these very moist blocks in the NoCP cases which explain their higher ensemble-mean rain production (measured in vertically-integrated rain water $\mathcal{R}$, fig. \ref{fig: bulk moisture and precip}g) and surface precipitation flux ($\mathcal{P}$, Fig. \ref{fig: bulk moisture and precip}g,h).

\subsection*{By reducing mesoscale ascents, cold pools arrest mesoscale self-aggregation of moisture}
\label{sec: 3}

\begin{figure}[t!]
\centering
\includegraphics[width =0.8\linewidth]{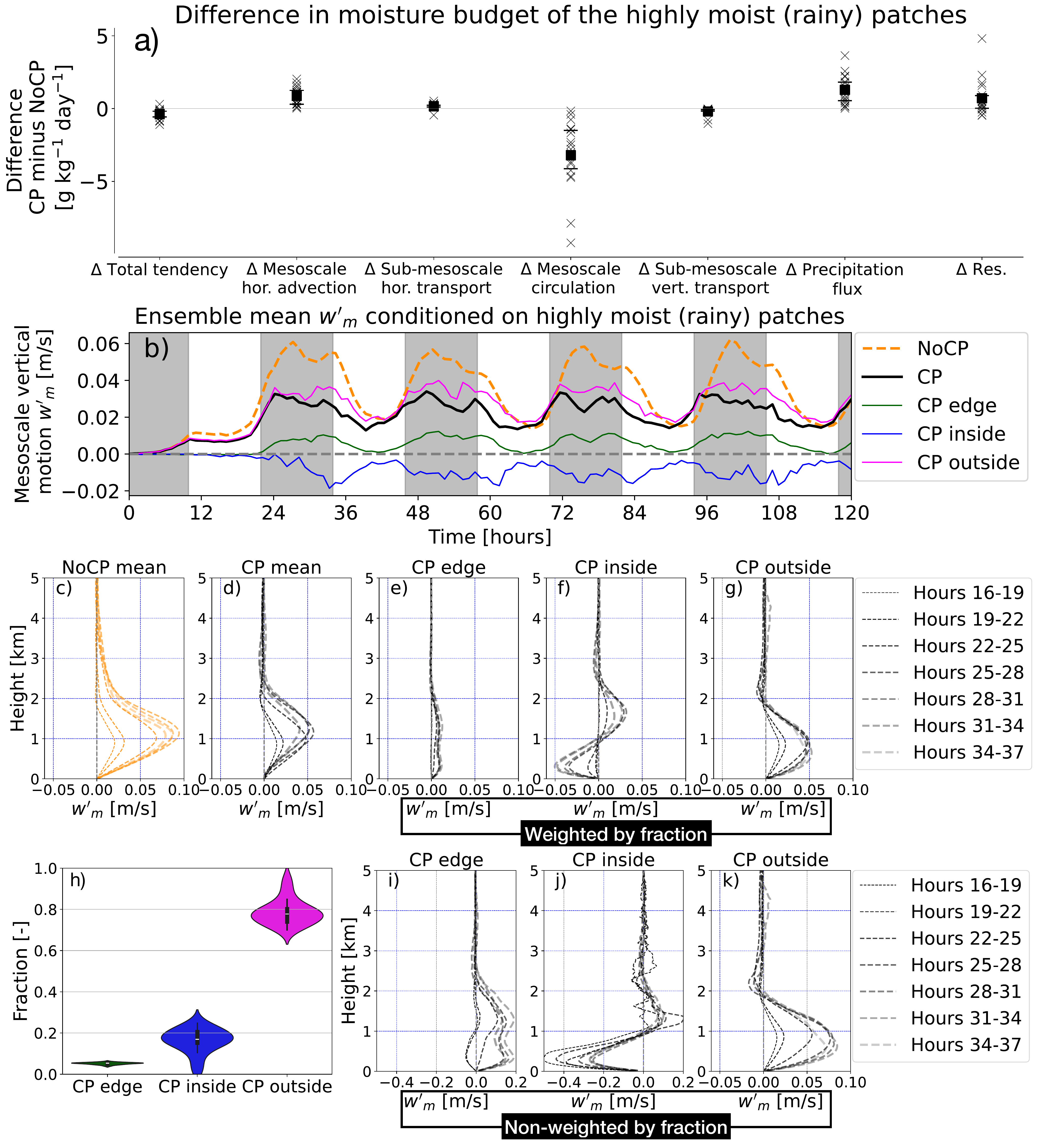}
\caption{{\textbf{Impact of cold pools on mesoscale self-aggregation of moisture.}} (a) Differences between terms in Eq. \ref{eq: final 10-km moisture} for the CP and NoCP ensembles. At each time step, components are averaged over 10-km blocks where $I'$ exceeds the 90th percentile. Time averages are taken over hours 12–36, when the simulations begin to diverge in total moisture. Crosses ($\times$) indicate individual simulation means, squares show the ensemble mean, and horizontal bars mark the interquartile range. (b) Ensemble-averaged time series of mesoscale 10-km vertical motion ($w'_m$) in the lowest 2 km for NoCP (orange) and CP (black), with CP contributions from cold-pool edges (green), inside (blue), and outside (magenta). (c-g) Vertical profiles of (b) for hours 16-37. (i-k) Same as (e-g), but without weighting by (h), which shows the distribution of the fractions of cold-pool regions within highly moist (rainy) 10-km blocks.}
\label{fig: moisture budget 10-km blocks}
\end{figure}

To understand why the CP ensemble lacks extremely moist mesoscale blocks, we perform another moisture budget analysis, but now on the 10-km blocks (see \textit{Methods}). In the CP ensemble, these blocks moisten at a rate 0.31 g/kg/day smaller than in the NoCP ensemble (Fig. \ref{fig: moisture budget 10-km blocks}a). This difference is mainly driven by weaker mesoscale circulations (more specific definition in \textit{Methods}) in the CP ensemble (Fig. \ref{fig: moisture budget 10-km blocks}a), which in turn is controlled by a relative weakening of mesoscale vertical ascents $w'_m$ (Fig. S2). The absence of very moist blocks in the CP ensemble is therefore primarily due to their weaker $w'_m$ compared to the NoCP ensemble (Fig. \ref{fig: moisture budget 10-km blocks}b).

To determine how cold pools reduce mesoscale ascent in the CP ensemble, we decompose $w'_m$ into contributions from three regions: inside cold pools, cold-pool edges, and outside cold pools (Fig. S3). We find that cold-pool interiors and edges exert opposing effects: strong downdrafts within cold-pool interiors, which cover larger areas, are offset by strong updrafts at cold-pool edges, which occupy much smaller fractions (Fig. \ref{fig: moisture budget 10-km blocks}d–k). Note that due to the structure of cold-pool–cloud coupling, part of the updrafts are sheared back over cold-pool interiors \cite{alinaghi2024external}, resulting in some updrafts occurring within cold pools at cloud level (Fig. \ref{fig: moisture budget 10-km blocks}f,j; see also Fig. S4). As a result of the offsetting between the cold-pool interiors and edges, the overall mesoscale ascent in the CP ensemble is governed by regions outside cold pools. On average, $w'_m$ outside cold pools in the CP ensemble evolves similarly to $w'_m$ in the NoCP ensemble (Fig. \ref{fig: moisture budget 10-km blocks}b,c,g,k). However, because cold pools, when averaged over time and space, occupy $\approx$20\% of these blocks and contribute nearly zero net $w'_m$ (Fig. \ref{fig: moisture budget 10-km blocks}h), the overall $w'_m$ in the CP ensemble is reduced relative to the NoCP ensemble.

\section{Summary and Discussion}
In summary, cold pools suppress mesoscale vertical ascent and arrest moisture aggregation into very moist mesoscale regions. This inhibition prevents the formation of strongly precipitating systems that drive a mean drying of shallow cumulus fields. By reducing this mean drying, cold pools limit the emission of longwave clear-sky radiation to space, leading to a net warming. The radiative warming of cold-pool dynamics (1.88\,W/m$^2$) is comparable to that induced by varying large-scale CCFs over a climatologically representative range.

These findings imply that mesoscale processes have the potential to  influence the simulated mean cooling of regimes covered by shallow cumulus clouds, depending on how models treat the interplay between cold-pool-driven and moisture-aggregation-driven circulations, and their effects on warm rainfall. Traditional GCMs do not represent either circulation, and would thus require other unresolved scales models (parameterizations) to produce the precipitation drying that in either of our model configurations is driven by the circulations. Conversely, emerging kilometer-scale GCMs \cite{stevens2019dyamond,satoh2019global}, which are capable of representing mesoscale self-aggregation \cite{saffin2023kilometer}, likely lack the resolution to resolve shallow cold pools \cite{fievet2023sensitivity}, and may thus require parameterizations that represent their effects to not develop a cold bias. Finally, LES models retain large uncertainties in their microphysics schemes, including rain evaporation, which is crudely represented even at resolutions as fine as 100\,m. Cold pools have been shown to be sensitive to microphysical parameterizations \cite{li2015sensitivity,kazil2025cold}. Hence, we find it plausible that mesoscale processes, which modulate the precipitation efficiency of shallow cumulus regimes, constitute an important overlooked process in GCMs and models of cloud-climate feedback.

Notably, the mesoscale cloud processes affect the mean cooling not through the clouds themselves, but through their organization of clear-sky water vapor. This is consistent with recent observational examinations of the subtropics \cite{fildier2023moiture}, but also with analyses of the tropical deep convective regime, which show that: (i) cold pools suppress convective self-aggregation \cite{jeevanjee2013convective,muller2013aggregation,boing2012influence}, and (ii) self-aggregation leads to mean drying of the tropics and enhances longwave clear-sky radiative cooling \cite{bony2020observed}. Therefore, we are now finding mechanisms across both tropical and subtropical regimes, through which mesoscale organization can shape clear-sky, longwave radiation. Hence, future work should explore if adding shallow mesoscale circulations to our theoretical frameworks, observational analyses, and global numerical models alters their estimated clear-sky longwave feedback.

\section{Materials and Methods}
\subsection*{Ensembles of large-eddy simulations}
\label{sec: methods ensemble desing}
To assess the impact of cold pools on top-of-atmosphere radiative budgets, we performed two 19-member ensembles of large-domain (153.6$\times$153.6\,km²) large-eddy simulations (LESs), with 100\,m horizontal and ~20\,m vertical resolution. Both ensembles are subsets of the \textit{Cloud Botany} ensemble \cite{jansson2023cloudbotany}, which is forced by climatologically relevant large-scale cloud-controlling factors (CCFs) of the trade-wind regime, deduced from ERA5 reanalysis data \cite{hersbach2020era5}. In this study, we select only those CCFs to which cold pools are sensitive (Table S1). Previous studies on the full ensemble \cite{alinaghi2024external,alinaghi2024cold} show that cold pools are strongly modulated by the geostrophic wind speed $\vert u_0 \vert$ and subsidence $w_1$, and to a lesser extent by free-tropospheric stability $\Gamma$ and wind shear $u_z$. As each CCF increases from weak to strong, cold-pool number and size increase from nearly zero to larger values, making this subset representative of the full ensemble. 

We could have added a fifth dimension of variability—one to which cold pools are not sensitive, such as free-tropospheric humidity—to our subset. In that case, this fifth CCF would need to be varied while keeping the other four CCFs (which strongly control cold pools) fixed at their values in the central reference simulation. As a result, cold-pool characteristics would remain similar across all simulations in that dimension. Consequently, the impact of homogenizing rain evaporation on the radiative budgets would match that of the central reference simulation. Therefore, including a fifth CCF to which cold pools are insensitive would not alter our estimates of the radiative impact of suppressing cold-pool formation across the ensemble.

\subsection*{Cold-pool formation inhibited by homogenizing rain-evaporation}
Cold pools are results of rain-evaporation-driven downdrafts. To understand their effect, we created two ensembles of LESs with similar CCFs: one with cold pools (CP) and one without (NoCP). In the NoCP cases, wherever in the simulation domain rain-evaporation starts to locally moisten and cool the atmosphere, this local tendency is evenly redistributed across the whole domain at each model level \cite{boing2012influence}. Homogenizing rain-evaporation in only the sub-cloud layer gives similar results.

\subsection*{Moisture budget analysis for the full domain}

We analyze moisture budget for the mass-weighted vertical average of total-water specific humidity $q_t$, defined as 
\begin{equation}
    I = \langle q_t \rangle =  \frac{\int_0^{z_\infty} \rho (z)q_t(z) dz}{\int_0^{z_\infty} \rho (z) dz},
\end{equation}
where $\rho (z)$ denotes the reference density profile, and $z_{\infty}$ is the domain top at 7 km. The variations of term $I$ is primarily driven by the boundary layer processes across the ensemble (see Fig. S5). Applying the mass-weighted averaging operator, defined as
\begin{equation}
    \langle \circ \rangle =  \frac{\int_0^{z_\infty} \rho (z)\circ dz}{\int_0^{z_\infty} \rho (z) dz},
\end{equation}
to all terms in the moisture budget equation, we obtain the bulk moisture budget:
\begin{equation}
\label{eq: qt budget}
\begin{split}
\frac{\partial I}{\partial t} = \underbrace{\frac{\partial \langle q_t \rangle}{\partial t}}_\text{Bulk moisture tendency} =\ 
&-\underbrace{\left\langle \frac{\partial \left(\overline{w'q'_t} \right)}{\partial z} \right\rangle}_{\text{Vertical flux convergence}} \\
&- \underbrace{ \left\langle u_j^{LS} \left( \frac{\partial q_t}{\partial x_j} \right)^{LS} \right\rangle}_{\text{Large-scale advection and subsidence}} \\
&- \underbrace{\left\langle P(z) \right\rangle}_{\text{Precipitation}} 
+ \underbrace{\langle N \rangle}_\text{Nudging},
\end{split}
\end{equation}
where $u_j^{LS}$ represents the large-scale velocity in the $j$-direction, $\overline{w'q'_t}$ is the slab-mean total moisture flux at height $z$ with 
\begin{equation}
    X' = X - \overline{X}, X \in \{w,q_t\},
\end{equation}
where $X'$ is anomaly of variable $X$ with respect to the slab-mean average $\overline{X}$. $P(z)$ is the domain-mean total precipitation flux at height $z$. Specifically, $P(z)$ is the sum of auto-conversion, accretion, and rain-evaporation fluxes. Note that the moisture profile in each member of the \textit{Cloud Botany} ensemble is nudged towards its initial horizontally-averaged value with the nudging term in Eq. \ref{eq: qt budget} \cite{jansson2023cloudbotany}.

\subsection*{Moisture budget analysis for the mesoscale blocks}

Similar analysis of Eq. \ref{eq: qt budget} can be written for the 10-km, mesoscale blocks; however, the mesoscale blocks also experience horizontal moisture fluxes at their boundaries with adjacent mesoscale blocks. Thus, the mesoscale blocks' moisture budget can be expressed as
\begin{equation}
\label{eq: lagrangian moisture budget}
\begin{split}
\frac{D I}{Dt} = \frac{D \langle q_t \rangle}{D t} =\ 
&- \underbrace{\left\langle \frac{\partial (U'_{h_j}q_t)}{\partial x_j} \right\rangle}_{\text{i}} 
- \underbrace{\left\langle \frac{\partial (w'q_t)}{\partial z} \right\rangle}_{\text{ii}} \\
&- \underbrace{ \left\langle u_j^{LS} \left( \frac{\partial q_t}{\partial x_j} \right)^{LS} \right\rangle}_{\text{Large-scale advection and subsidence}} \\
&- \underbrace{\left\langle P(z) \right\rangle}_{\text{Precipitation}} 
+ \underbrace{\langle N \rangle}_\text{Nudging},
\end{split}
\end{equation}
where the first ($\text{i}$) and second ($\text{ii}$) terms on the right-hand side represent the total $q_t$ flux in the horizontal and vertical, respectively. Here, $U'_{h_j}$ and $w'$ denote the horizontal and vertical components of the anomalous velocity vector with respect to the domain mean. This means that the horizontal advection with the domain-mean horizontal wind is absorbed in the left-hand side of Eq. \ref{eq: lagrangian moisture budget}, which means our budget analysis follows the mesoscale blocks as they move with the larger-scale wind.

To understand the contributions from the mesoscales and sub-mesoscales, we decompose variables $\phi \in \{U'_{h_j},w'\}$ to contributions from the mesoscales $\phi_m$ and sub-mesoscales $\phi_s$. The mesoscale contributions $\phi_m$ are calculated by averaging $\phi$ over each 10-km block. For each 100-m grid cell inside a 10-km block, sub-mesoscale contributions $\phi_s$ is anything that remains after filtering the mesoscale contributions, i.e. $\phi_s = \phi - \phi_m$. By applying the product rule to the first two terms ($\text{i}$ and $\text{ii}$) of Eq. \ref{eq: lagrangian moisture budget}, we obtain
\begin{equation}
    \langle\text{i}_m\rangle = -\left\langle U'_{h_{j,m}}\frac{\partial q_{t,m}}{\partial x_j} \right\rangle - \left\langle q_{t,m} \frac{\partial U'_{h_{j,m}}}{\partial x_j} \right\rangle -\left\langle \frac{\partial (U'_{h_{j,s}}q_{t,s})_m}{\partial x_j} \right\rangle
    ,
\end{equation}
\begin{equation}
     \langle\text{ii}_m\rangle = 
    - \left\langle w'_{m} \frac{\partial q_{t,m}}{\partial z} \right\rangle - \left\langle q_{t,m} \frac{\partial w'_{m}}{\partial z} \right\rangle - \left\langle\frac{\partial (w'_s q'_{t,s})_m}{\partial z} \right\rangle.
\end{equation}
The mass conservation at the mesoscales implies that $q_{t,m} \frac{\partial U'_{h_{j,m}}}{\partial x_j} + q_{t,m} \frac{\partial w'_{m}}{\partial z} = 0$. Substituting this relation, we obtain
\begin{equation}
\label{eq: final 10-km moisture budget}
\begin{split}
\langle\text{i}_m\rangle + \langle\text{ii}_m\rangle =\ 
&\underbrace{-\left\langle U'_{h_{j,m}}\frac{\partial q_{t,m}}{\partial x_j} \right\rangle}_{\text{mesoscale horizontal advection}} \\
& \underbrace{- \left\langle \frac{\partial (U'_{h_{j,s}}q_{t,s})_m}{\partial x_j}\right\rangle}_{\text{sub-mesoscale horizontal transport}} \\
&\underbrace{- \left\langle w'_{m} \frac{\partial q_{t,m}}{\partial z}\right\rangle}_{\text{gradient production}}  
\underbrace{- \left\langle\frac{\partial (w'_s q'_{t,s})_m}{\partial z}\right\rangle}_{\text{sub-mesoscale vertical transport}}.
\end{split}
\end{equation}

In Fig.\,\ref{fig: moisture budget 10-km blocks}a, we refer to the term “gradient production” as “mesoscale circulation” to make it more accessible to a general reader. Because the 10-km blocks in both ensembles experience the same external large-scale subsidence and advection and are similarly nudged towards their initial state, we include both terms in the residual and rewrite Eq. \ref{eq: lagrangian moisture budget} as
\begin{equation}
\label{eq: final 10-km moisture}
\begin{split}
    \frac{D I_m}{Dt} = \frac{D \langle q_{t,m} \rangle}{Dt} = \langle\text{i}_m\rangle + \langle\text{ii}_m\rangle -\langle P_m(z) \rangle + \text{Residual}_m.
\end{split}
\end{equation}
To analyze moisture aggregation in the very moist blocks of both ensembles, we calculate the terms of Eq. \ref{eq: final 10-km moisture} for these blocks at each hour for each ensemble member. At every time step, the terms are averaged over blocks where mesoscale moisture anomaly $I'_m$ exceeds the 90$^{th}$ percentile of its distribution for that time step. This condition closely resembles averaging over blocks where the rain-water path exceeds the 90$^{th}$ percentile of its distribution. The differences in the terms of Eq. \ref{eq: final 10-km moisture} between the CP and NoCP ensembles are shown in Fig. \ref{fig: moisture budget 10-km blocks}a. This comparison helps identify which term is responsible for the higher humidity in the very moist blocks of the NoCP ensemble compared to their counterparts in the CP ensemble.

\section*{Data Availability}
Datasets from the large-eddy simulation ensembles, along with the Python scripts used to generate the figures in this manuscript, are publicly available by \cite{alinaghi_2025_15544026} at \url{https://doi.org/10.5281/zenodo.15544026}. The version of the Dutch Atmospheric Large-Eddy Simulation for homogenizing rain-evaporation is publicly available at \url{https://github.com/dalesteam/dales/tree/botany-homogenization}.

\section*{Acknowledgments}
PA deeply thanks Steven Boing for his help on homogenizing rain-evaporation in DALES. During the time working on this research, PA received support from The Branco Weiss Fellowship - Society in Science, administered by ETH Zurich. PA and MJ acknowledge support from the International Space Science Institute (ISSI) in Bern, through ISSI International Team project 576 (“Constraining Trade‐Cumuli Feedback by Means of Process Understanding”). This research used computational resources of Fugaku provided by RIKEN through the HPCI System Research Project (Project ID: hp240116).

\bibliographystyle{apacite}
\bibliography{citations}

\newpage
\appendix
\section*{Supplementary Information}
This part contains a table and figures to additionally support the text in the main document.

\setcounter{figure}{0} 
\renewcommand{\thefigure}{S\arabic{figure}}

% \subsection*{Subhead}
% Type or paste text here. This should be additional explanatory text such as an extended technical description of results, full details of mathematical models, etc.   

% \section*{Heading}
% \subsection*{Subhead}
% Type or paste text here. You may break this section up into subheads as needed (e.g., one section on ``Materials'' and one on ``Methods'').

% \subsection*{Materials}
% Add a materials subsection if you need to.

% \subsection*{Methods}
% Add a methods subsection if you need to.

%%% Each figure should be on its own page
\begin{figure}
\centering
\includegraphics[width=\textwidth]{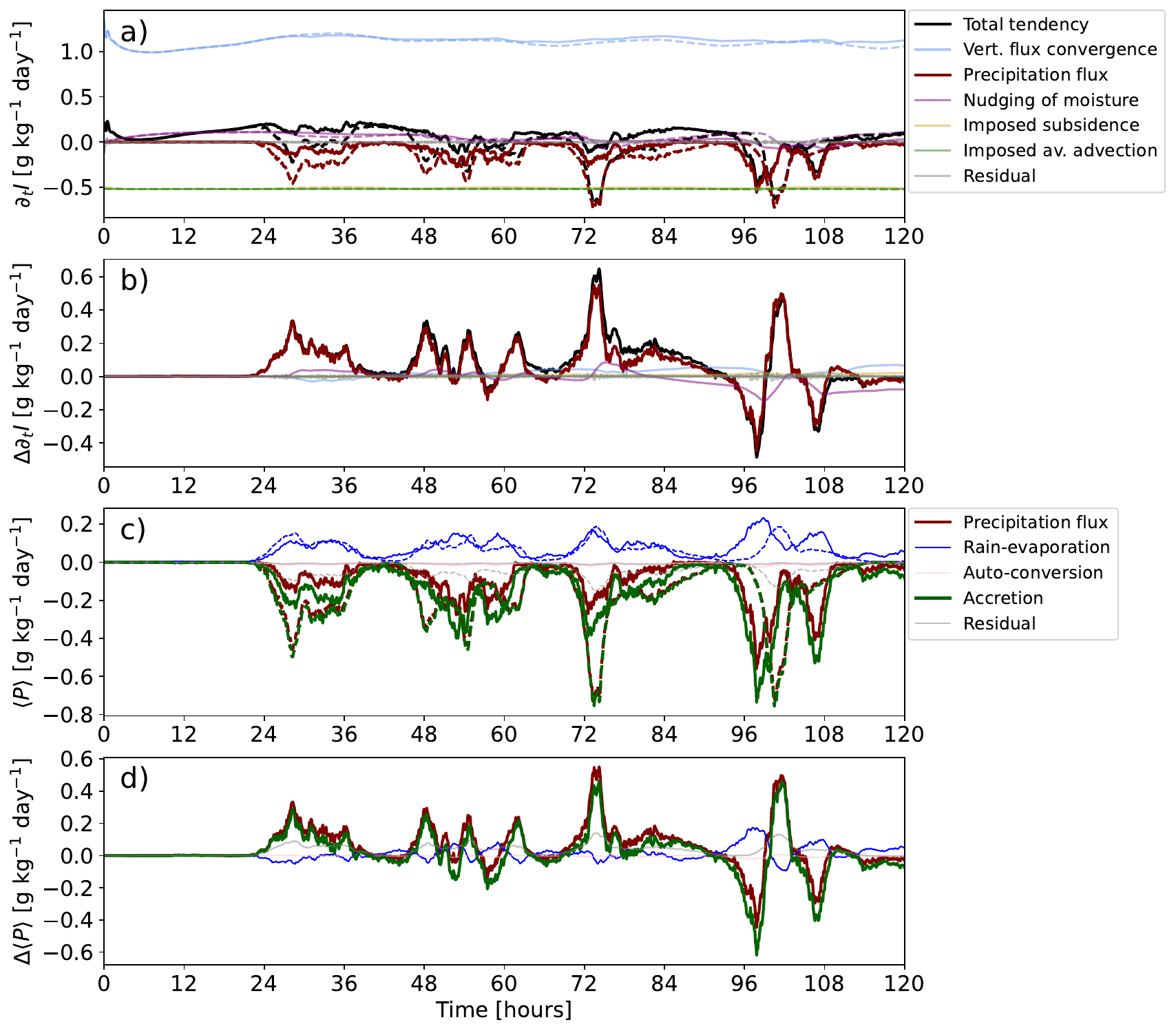}
\caption{\textbf{Bulk moisture budget time series for the central reference simulation.} Time series of (a) the components of the bulk moisture budget for the CP (solid lines) and NoCP (dashed lines) ensembles, and (b) their difference (CP minus NoCP). Panels (c) shows the rain-related components of the moisture budget again for the CP (solid) and NoCP (dashed) ensembles, with their difference shown in (d).}
\label{fig: bulk moisture budget}
\end{figure}

\begin{figure}
\centering
\includegraphics[width=\textwidth]{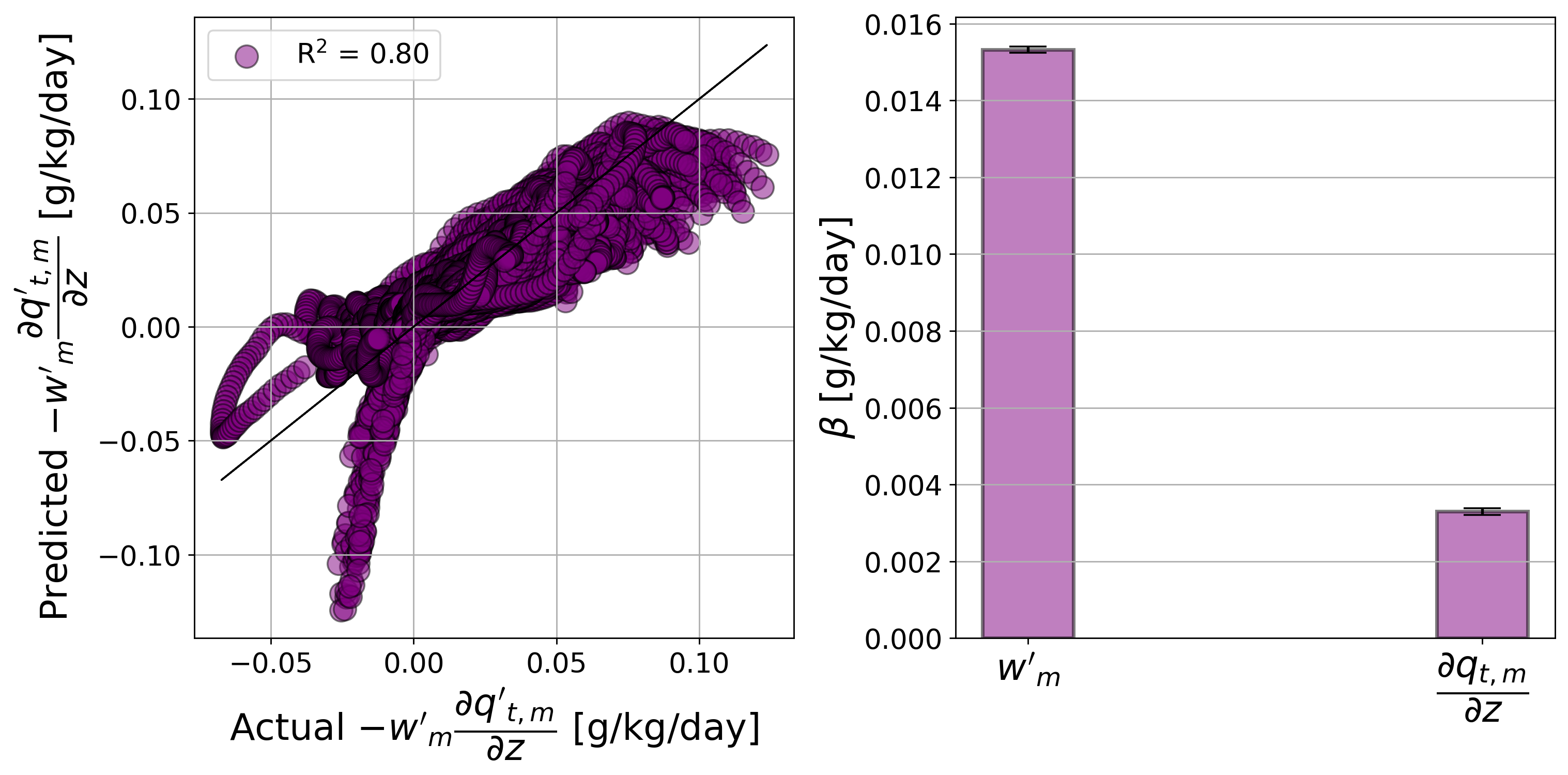}
\caption{\textbf{Multivariate regression results of the gradient production term $-w'_{m} \dfrac{\partial q_{t,m}}{\partial z}$.} (Left) The scatter plot of the results of multivariate regression analysis with around 130 k points from the mesoscale blocks of the entire ensemble in 3D. (Right) The standardized beta coefficients of the multiple regression analysis for predicting the gradient production term ($-w'_{m} \dfrac{\partial q_{t,m}}{\partial z}$) as a function of mesoscale vertical ascent $w'_m$ and gradient of mesoscale moisture $\dfrac{\partial q_{t,m}}{\partial z}$. The error bars show the confidence intervals at the 99.99$^{th}$ level of confidence. \textbf{Message} is that the gradient production term is strongly controlled by the mesoscale vertical motion and the effect of $w'_m$ is around five times larger than that of moisture gradient in height.}
\label{fig: regression gradient production}
\end{figure}

\begin{figure}
\centering
\includegraphics[width=\textwidth]{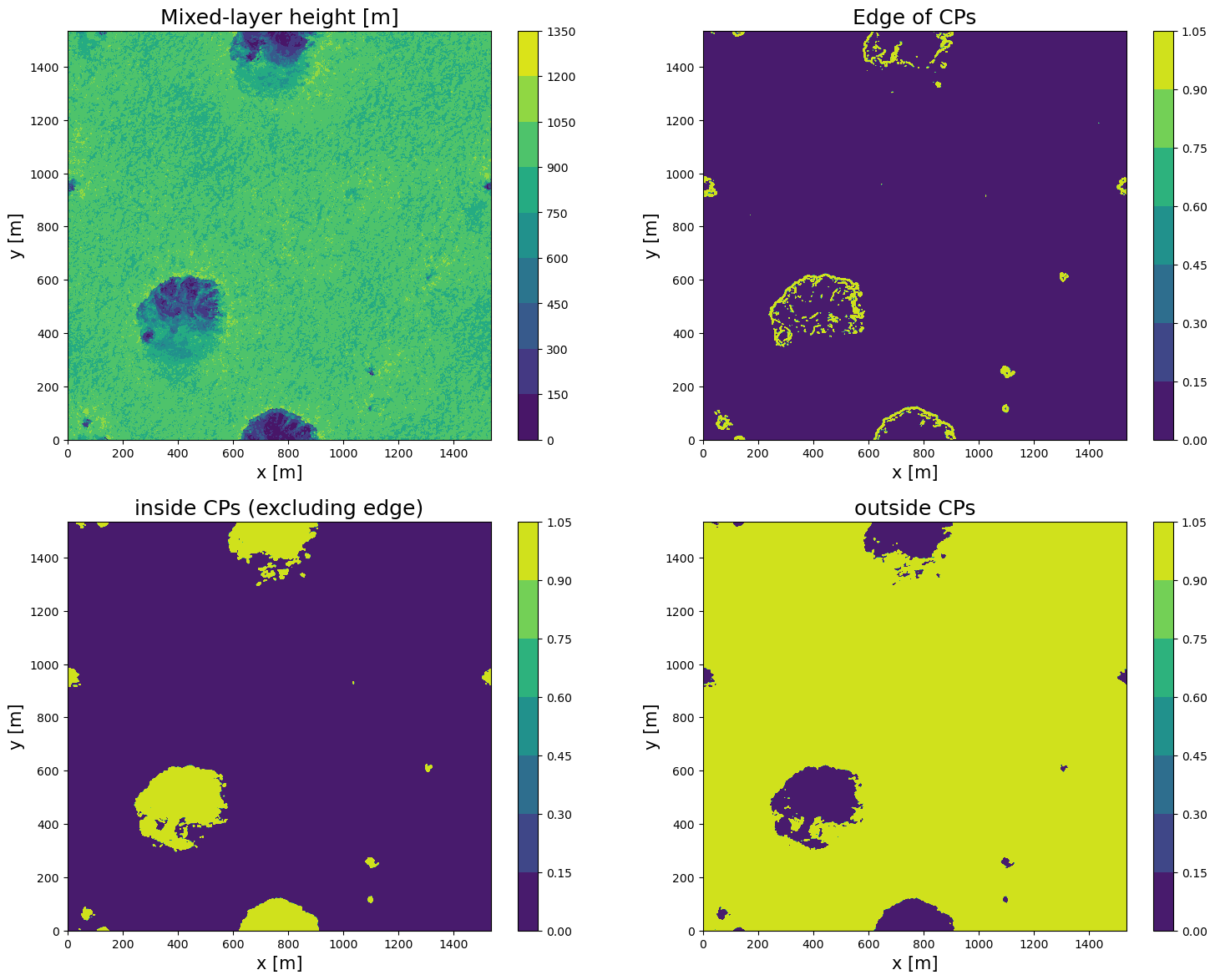}
\caption{\textbf{An example of identifying regions inside, at the edge, and outside cold pools.} We identify cold pools following Alinaghi et al. (2025, \textit{External drivers and mesoscale self-organization of shallow cold pools in the trade-wind regime}). The edge of cold pools is determined by identifying the regions where there is a sharp change in the mixed-layer height. Regions \textit{inside} cold pools exclude the regions at the edge. \textit{Outside} the cold pools refers to all areas in the simulation domain that are neither inside nor at the edge of cold pools. For complete details, see the Python codes provided at the link in the supplementary materials.}
\label{fig: cold pools}
\end{figure}

\begin{figure}
\centering
\includegraphics[width=\textwidth]{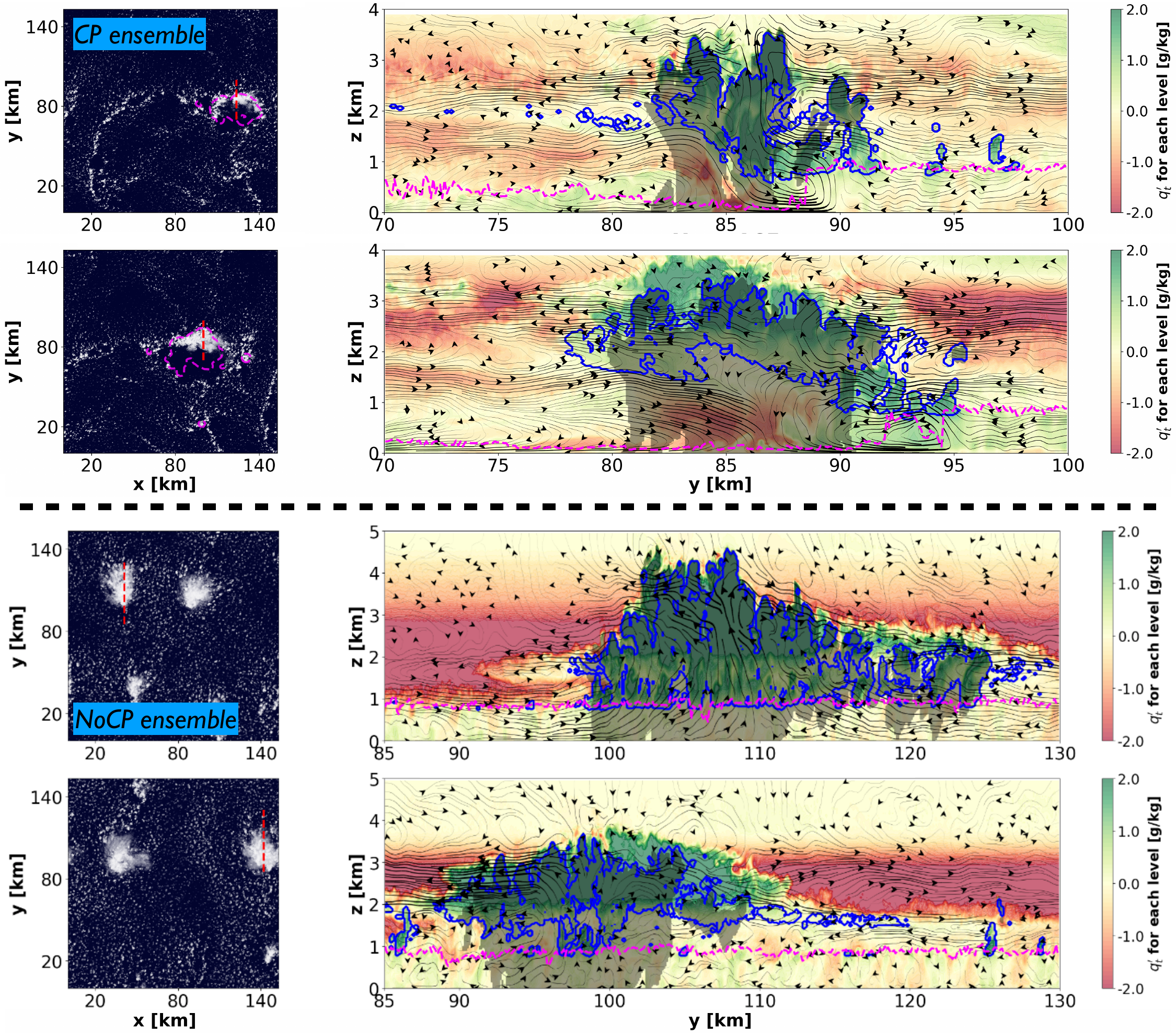}
\caption{\textbf{Cloud-circulation coupling in CP vs. NoCP ensembles.} The left column shows four contour plots of cloud albedo along with the cold-pool mask shown by the dashed, magenta contour lines for two time steps of CP (upper part) and NoCP (lower part) ensemble. The right column shows y-z-cross-section contour plots of the total specific humidity anomalies associated with the dashed red line in snapshots shown in the left column. Cloud and rain boundaries are shown by the blue and grey contour lines. Circulations are made from the meridional ($v$) and vertical ($w$) velocity anomalies and are shown by black streamlines. To reduce the noise from the circulations, $v,w$ are made from the medians of a 5-km window along the x dimension. The $h_{mix}$ is shown by the dashed, magenta contour lines. The South (S), North (N), West (W), and East (E) directions are shown by orange labels.}
    \label{fig: cloud circ coupling}
\end{figure}

\begin{figure}
\centering
\includegraphics[width=0.7\textwidth]{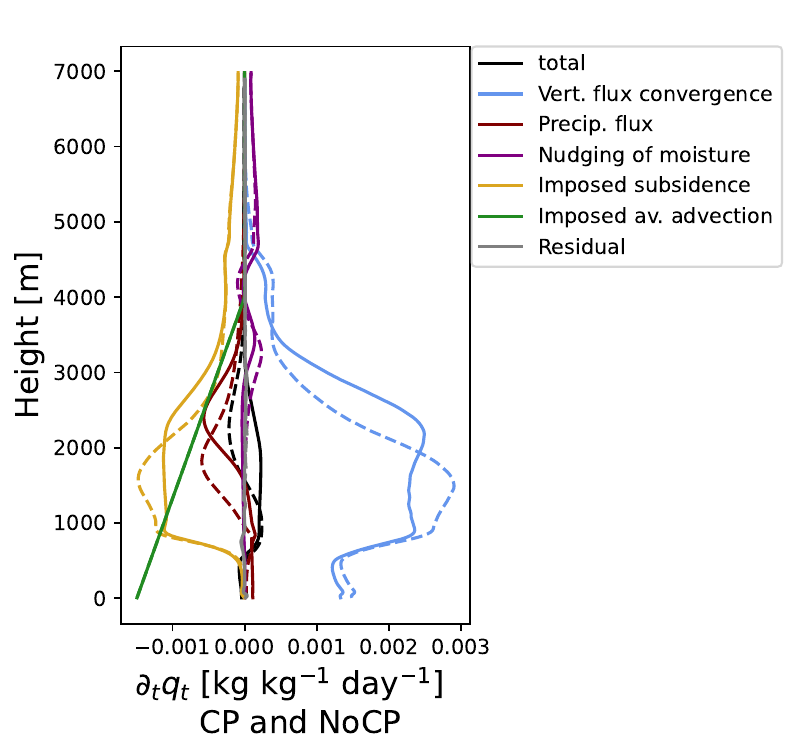}
\caption{\textbf{Vertical profiles of the moisture budget components for the central reference simulation.} This plot shows the vertical slab-mean profiles of the components of the moisture budget shown in Fig.\,\ref{fig: bulk moisture budget}a, for the central reference simulations of the CP (continuous) and NoCP (dashed) ensembles, averaged over hours 0-120. The figure highlights that the variations of total-water specific humidity $q_t$ and in turn the mass-weighted vertically-averaged moisture $I$ (introduced in \textit{Methods}) in our simulations are primarily driven by the boundary-layer processes.}
\label{fig: moisture budget profiles}
\end{figure}

\begin{table}[ht!]
\begin{tabular}{|| m{7cm} | m{3.5cm} | m{2cm} | m{2cm} ||} 
 \hline
 \textbf{Large-scale and
    initial conditions} & \textbf{Parameters [units]} & \textbf{Number of simulations} &  \textbf{Range of variability}\\ [0.5ex] 
 \hline
Near-surface geostrophic wind speed & $u_0$ [m/s] & 5 &from 5 to 15\\ 
\hline
Temperature lapse rate in the free troposphere & $\Gamma$ [K/km] & 5 & from 4.5 to 7.5\\ 
 \hline
Large-scale vertical velocity variability & $w_1$ [cm/s] & 4 & from $-0.002$ to 0.001\\
\hline
Shear in the horizontal geostrophic wind & $u_z$ [(m/s)/km] & 4 & from $-4$ to 4\\
 \hline
\end{tabular}
\caption{\textbf{Parameters of the LES ensemble.} Overall information about the ensemble's parameters determining CCFs. Note that in addition to simulations above, the ensemble has a central reference simulation with the mean of CCFs above. This means our ensemble features 19 simulations in total.}
\label{tab1: CCFs params}
\end{table}

\end{document}